# Generalized effective mass approach for cubic semiconductor n-MOSFETs on arbitrarily oriented wafers

Anisur Rahman*, Mark S. Lundstrom, and Avik W. Ghosh

School of Electrical and Computer Engineering

1285 Electrical Engineering Building

Purdue University, West Lafayette, IN 47907

## Abstract

The general theory for quantum simulation of cubic semiconductor n-MOSFETs is presented within the effective mass equation approach. The full three-dimensional transport problem is described in terms of coupled transverse subband modes which arise due to quantum confinement along the body thickness direction. Couplings among the subbands are generated for two reasons: due to spatial variations of the confinement potential along the transport direction, and due to non-alignment of the device coordinate system with the principal axes of the constant energy conduction band ellipsoids. The problem simplifies considerably if the electrostatic potential is separable along transport and confinement directions, and further if the potential variations along the transport direction are slow enough to prevent dipolar coupling (Zener tunneling) between subbands. In this limit, the transport problem can be solved by employing two unitary operators to transform an arbitrarily oriented constant energy ellipsoid into a regular ellipsoid with principal axes along the transport, width and confinement directions of the device. The effective masses for several technologically important wafer orientations for silicon and germanium are calculated in this paper.

* rahmana@purdue.edu, phone: 765-494-9034, fax: 765-494-6441

## I. Introduction

Metal oxide semiconductor field effect transistors (MOSFETs) constitute the fundamental building block of present day CMOS technology. Current research in this field is largely geared towards improving MOSFET performance and increasing device density through aggressive scaling of their feature sizes [1]. The importance of quantum mechanical size effects in MOSFETs, where the inversion layers are just a few nanometers thick, was realized during the early period of their development [2, 3]. Moreover, as the channel length of the MOSFETs approaches few tens of nanometers, source-to-drain and gate tunneling in these near-ballistic devices become important issues [4]. Numerical simulations provide valuable insight into the physics of device operation at this scale, requiring an appropriate treatment of the device bandstructure as well as a rigorous formulation of quantum transport.

The effective mass equation (EME) provides an accurate, easy to implement model Hamiltonian that does justice to the device bandstructure including quantum confinement effects within the inversion layer, and describes the slowly varying envelope part of the underlying Bloch wavefunction. The Non-Equilibrium Green's Function (NEGF) method provides a rigorous formulation of quantum transport in nanoscale devices [5]. Together, the NEGF formalism and the effective mass equation have been used to describe transport in nanoscale MOSFETs both in the ballistic limit [6-9], as well including the effects of carrier scattering [10, 11]. In [7] and [12] the coupled- and decoupled-mode-space approaches were introduced, and in [13] the coupled-mode-space approach is used in order to assess the effects of channel access geometry and series resistance in nanoscale n-MOSFETs. In the presence of strong dephasing with bandlike transport, the NEGF equation reduces to the semiclassical Boltzmann transport equation (BTE). The BTE has also been used, along with related concepts such as density-of-states and conduction effective masses, in order to explore the upper limit of nanoscale MOSFET performance [14-16]. The 2D numerical simulator nanoMOS 2.5, a freeware, has been developed to simulate the ballistic and scattering characteristics of ultra-thin-body (UTB), double gate (DG) n-MOSFETs using both semiclassical (BTE) and fully quantum (NEGF) methods [17, 18].

Silicon (100) wafers are almost universally used by the semiconductor industry for CMOS integrated circuit fabrication. Simulation of n-MOSFETs is generally performed for devices fabricated on (100) wafers, motivated by its technological importance. The quantum simulation of Si (100) devices is substantially simplified by the fact that the principal axes of the six fold degenerate conduction band ellipsoids are aligned along the device coordinate axes, effectively decoupling the kinetic energies along the device coordinate axes. In general, however, the principal axes of the conduction band ellipsoids are not aligned with the device axes, so that the associated kinetic energies become coupled and the effective mass equation (EME) becomes nontrivial. Such a situation arises for transistors that employ germanium as a high-mobility channel material [19-21], as well as for alternate wafer orientations of silicon [22]. To extend the application of EME to analyze these novel n-MOSFETs it is necessary to generalize the EME approach to arbitrary wafer orientations. In the past, Stern *et al.* proposed a method which, in such non-trivial cases, decouples the kinetic energy associated with the quantum confinement direction from that associated with the motion in the transport plane [2], [23]. In this paper we introduce a technique that decouples the energy associated with all three device axes, i.e., transport, width and confinement directions, for devices with unvarying cross sections and slowly varying channel directed potentials. This allows us to use all the EME based simulation tools developed so far for modeling novel channel material n-MOSFETs.

This paper is organized as follows. In Sec. II we outline our general solution procedure, describing the full 3-D problem, and the conditions under which it can be simplified. In Sec. III we discuss the conduction band structure in cubic semiconductors and derive the effective mass tensor (EMT) in an arbitrary, orthogonal device coordinate system. In Sec. IV we present the technique to solve the resulting EME for n-MOSFETs. This general technique shows that under certain conditions one can employ two unitary transformations that map any arbitrarily oriented constant energy ellipsoid onto a regular ellipsoid having principal axes oriented along the device coordinate axes. In Sec. V, the effective masses are calculated for several technologically significant silicon and germanium wafer orientations. We follow by presenting some results and discussions in Sec. VI. Finally, we conclude in Sec. VII.

## II. Summary of The OVERALL Solution Procedure

The complete problem involves the full three dimensional quantum transport and electrostatics of the system. We start by writing down the non-diagonal EMT in the device coordinate system, and the corresponding dispersion relationship (11) for the arbitrarily oriented conduction band ellipsoids. We then perform a basis transformation which recasts the general effective mass equation in a *fully equivalent* form in terms of a complete set of transverse subband eigenmodes. These transverse modes are obtained by considering the confinement potential along the principal axis directions of the constant energy ellipsoids (18). The general three dimensional nature of the problem is manifested in (29) through couplings among these subbands, so that the corresponding coupled-mode-space transport formulation is fully equivalent to the original effective mass equation (13), *with no further assumptions*. Some of the coupling terms disappear if the confinement potential is unvarying along the transport direction, making the overall electrostatic potential separable (30a-b); at this stage, however, there still are couplings among the different subbands representing subband-to-subband Zener tunneling, caused by the non-alignment of the ellipsoidal principal axes with the device coordinate axes. These couplings are *not* present when the device axes coincide with the ellipsoidal axes, as in Si (100) devices. The origin of these couplings can be traced back to the non-alignment of the axes, which complicates the description of kinetic energy in the device coordinate system through the non-diagonal effective mass tensor (11). In effect the couplings arise because the channel potential itself is varying along the confinement direction, effectively coupling the two coordinates. If now the transport potential varies slowly enough that the total variation in channel potential between the confinement planes along the ellipsoidal axis is much smaller than the subband separation (33), the intersubband coupling terms are further eliminated, leading to a simplified decoupled mode-space description in terms of isolated ellipsoids with their principal axes oriented along the device axes (35), albeit with modified effective masses. The problem is further simplified computationally for ultra-thin body MOSFETs with a large energy difference between the transverse subbands, so that only the lowest few modes that are thermally populated need to be considered.

## III. CONDUCTION BAND STRUCTURE IN CUBIC SEMICONDUCTORS

The conduction band (CB) minima of cubic semiconductor materials appear either at a single point (for direct bandgap materials such as GaAs) or at multiple equivalent points (for indirect bandgap materials e.g. Si, Ge) within the first Brillouin zone (BZ). The constant energy surfaces become non-parabolic and warped for energies away from the band minima; close to the band edges, however, the relevant electronic states for transport calculations can be described by simple ellipsoidal surfaces. We can safely ignore the coupling with the valence band for semiconductors with moderately large bandgaps. Under these circumstances, the constant energy surface for electrons in a direct bandgap material is spherical, centered on the Γ-point and described as

$$E = \frac{\hbar^2 k^2}{2m_{eff}}, \tag{1}$$

with a constant, isotropic effective mass, $m_{eff}$. For indirect semiconductor materials, the CB minima are located at multiple equivalent points: six points near $X$ along the $\Delta$ or $\langle 100 \rangle$ crystallographic directions for silicon, and eight equivalent points at $L$ along $\Lambda$ or $\langle 111 \rangle$ for germanium. In indirect semiconductors, the constant energy surfaces are ellipsoids of revolution around $\Delta$ and $\Lambda$ axes, respectively [24, 25], requiring two effective masses, longitudinal $m_l$, and transverse $m_t$, for description. In general the non-alignment of the ellipsoidal principal axes with the device coordinate axes causes the effective mass to become a $3 \times 3$ tensor quantity in the device coordinate system [26, 27]. In this section, we will systematically derive this effective mass tensor (EMT) in an arbitrary orthogonal coordinate system, and in the subsequent section we will simplify the resulting EME for quantum transport simulation of n-MOSFETs.

We formulate the generalized EME by defining three orthogonal coordinate systems, presented schematically in Fig. 1. They are called: the device coordinate system (DCS), the crystal coordinate system (CCS), and the ellipsoid coordinate system (ECS). Three unit vectors, $\hat{k}_1$, $\hat{k}_2$ and $\hat{k}_3$ span the DCS and form its basis. We take $\hat{k}_3$ along the body thickness (i.e. quantum confinement of inversion carriers), $\hat{k}_1$ along the source to drain (i.e. transport) direction, and $\hat{k}_2 \left( \equiv \hat{k}_3 \times \hat{k}_1 \right)$ along the device width direction. The second coordinate system, CCS, is spanned by three unit vectors $\hat{k}'_1$, $\hat{k}'_2$ and $\hat{k}'_3$, oriented along the three orthogonal $\langle 100 \rangle$ crystallographic directions of the underlying channel material. Finally, the basis for the ECS consists of the unit vectors $\hat{k}_{\parallel}$, $\hat{k}_{\perp 1}$ and $\hat{k}_{\perp 2}$, chosen along the principal axes of each constant energy ellipsoid. In summary, the CCS is unique for all our simulations, the DCS depends on the fabrication choice (that is, on the wafer orientation and the source-to-drain direction in the chip design layout) and the ECS depends on the specific channel material and is unique to each ellipsoid.

We now describe the key steps in determining the EMT for a general conduction band ellipsoid. In the ellipsoid coordinate system (ECS) the constant energy ellipsoid can be expressed as:

$$E = \frac{\hbar^2 k_{\parallel}^2}{2m_l} + \frac{\hbar^2 \left( k_{\perp 1}^2 + k_{\perp 2}^2 \right)}{2m_t}. \tag{2}$$

In (2), the *k*-space origin is translated to the conduction band minima, which serves as the reference for the electronic energy. In compact vector notation, (2) can be written as

$$E = \frac{\hbar^2}{2} \vec{k}_E^T \left[ M_E^{-1} \right] \vec{k}_E \tag{3}$$

where $\vec{k}_E = \left( k_{\parallel}\, k_{\perp 1}\, k_{\perp 2} \right)^T$ consists of the component of an arbitrary wave vector in the ECS, and the inverse EMT, $\left[ M_E^{-1} \right]$, is a $3 \times 3$ diagonal matrix with $m_l^{-1}$, $m_t^{-1}$ and $m_t^{-1}$ along the diagonal. For a given channel material and for a given conduction band ellipsoid, the directions of the unit basis vectors $\hat{k}_{\parallel}$, $\hat{k}_{\perp 1}$ and $\hat{k}_{\perp 2}$ relative to the CCS are known, thus allowing us to write the $3 \times 3$ rotation matrix $\Re_{E \leftarrow C}$, which transforms the components of an arbitrary vector $\vec{k}_C \equiv \left( k'_1\, k'_2\, k'_3 \right)^T$ defined in the CCS, to its components in the ECS,

$$\vec{k}_E = \Re_{E \leftarrow C}\, \vec{k}_C\,. \tag{4}$$

A similar rotation matrix $\Re_{C\leftarrow D}$ transforms a wave vector $\vec{k}_D \equiv (k_1\, k_2\, k_3)^T$ in the DCS to $\vec{k}_C$ in CCS as

$$\vec{k}_C = \Re_{C\leftarrow D}\,\vec{k}_D\,. \tag{5}$$

Combining (4) and (5) we obtain,

$$\vec{k}_E = \Re_{E\leftarrow D}\,\vec{k}_D\,, \tag{6}$$

where the rotation matrix is defined as

$$\Re_{E\leftarrow D} = \Re_{E\leftarrow C}\Re_{C\leftarrow D}\,. \tag{7}$$

Inserting (6) into (3) we obtain

$$E = \frac{\hbar^2}{2}\vec{k}_D^T\left[M_D^{-1}\right]\vec{k}_D\,, \tag{8}$$

where the inverse effective mass $\left[M_D^{-1}\right]$ in the DCS is

$$\left[M_D^{-1}\right] = \Re_{E\leftarrow D}{}^T\left[M_E^{-1}\right]\Re_{E\leftarrow D}\,. \tag{9}$$

In Sec. V we will evaluate $\Re_{E\leftarrow D}$ for various wafer orientations and for both $\Delta$ and $\Lambda$ type CB valleys. From (9) we find that the general EMT, $\left[M_D^{-1}\right]$, is a full $3\times 3$ symmetric matrix whose elements $\left[M_D^{-1}\right]_{ij}$ are

$$\frac{1}{m_{ij}} = \frac{a_{1i}a_{1j}}{m_l} + \frac{a_{2i}a_{2j} + a_{3i}a_{3j}}{m_t}\ , \tag{10}$$

where $a_{ij} = \left[\Re_{E\leftarrow D}\right]_{ij}$. Equation (8) can now be written in compact form as

$$E(k_1,k_2,k_3) = \sum_{i,j=1}^{3}\frac{\hbar^2 k_i k_j}{2m_{ij}},$$

From (10) we find $m_{ij} = m_{ji}$ and therefore the above expression can be rewritten as

$$E(k_1, k_2, k_3) = \sum_{i=1}^{3} \frac{\hbar^2 k_i^2}{2m_{ii}} + 2\sum_{i=1}^{3} \sum_{i<j\leq 3} \frac{\hbar^2 k_i k_j}{2m_{ij}}. \quad (11)$$

Comparing with (1), we see that the expression for the constant energy ellipsoidal surface in (11) contains additional cross terms $k_i k_j$ in kinetic energy. In the following section we will see that this makes the corresponding general EME nontrivial and there we will outline our treatment of the problem.

## IV. THE GENERALIZED EME AND THE SOLUTION

The bulk bandstructure for any semiconductor is calculated by solving Schrödinger's equation using Bloch's theorem for periodic lattice. Although this technique yields an accurate description of the $E-\vec{k}$ relationship for the electrons over the entire BZ, it is unnecessarily complicated for treating transport problems in the MOSFET device structure. Since an accurate description of only the band edge electronic states is sufficient for transport simulation, the effective mass approximation scheme (also known as the envelope function approximation) becomes an attractive alternative. The effective mass approximation uses an accurate description of the $E-\vec{k}$ relationship over only a limited range of energy near the valence or conduction band extrema. The $\vec{k}\bullet\vec{p}$ perturbation technique is employed in this regard, which describes the bandstructure over a limited range of energy near the band edge with sufficient accuracy [28], and the Schrödinger like effective mass equation (EME) is obtained by replacing certain components of the wave vector, $k_j$, in the expression $E(\vec{k})$ with their quantum mechanical operator, $-i\frac{\partial}{\partial x_j}$, and the electronic states are obtained by solving this differential eigenvalue equation.

The EME scheme described above is universally used by the electronic device community for quantum mechanical simulation of MOSFETs. For silicon devices fabricated on (100) wafers with the source to drain direction oriented along [010], the EMT continues to be diagonal, and therefore, the cross terms in (11) drop out. In this case, the EME can be solved without difficulty and the quantum mechanical effects are accurately included in the simulation. Since the pioneering work by Stern [3], the above mentioned device orientation has been exclusively used for simulation and to the knowledge of the authors, no work has been done to generalize the EME approach for n-MOSFETs fabricated on arbitrarily oriented wafers with a general source-to-drain direction. As a result, the quantum mechanical simulation of MOSFETs fabricated on germanium (001) wafers or silicon (111) or (110) wafers still remains a nontrivial problem, since their effective mass tensors in the device coordinate systems are full $3\times3$ matrices. In this section we will introduce the general solution technique and will show that under certain simplifying conditions we can decouple the energies in (11), thereby, eliminating the limitations of EME stated above.

In Fig 2 the ultra-thin-body double-gate SOI MOSFET device structure is shown as the model device; however, the general theory we are developing is valid for bulk MOSFETs as well. In this figure we have defined $X$, $Y$ and $Z$ as the real space Cartesian axes along the previously mentioned $\hat{k}_1$, $\hat{k}_2$ and $\hat{k}_3$ unit vectors, respectively. Accordingly, we replace the *k*-subscripts in (11) from {*123*} to {*xyz*}. For the MOSFET, $X$, $Y$ and $Z$ represent the transport, quantum confinement and width directions, respectively.

Our general strategy will be to first solve the quantum problem along the confinement direction and then use the corresponding eigenvectors to construct the complete basis set for the full three dimensional quantum transport problem. Using the new labels for the axes of the DCS, described above, the general $E(\vec{k})$ relation in (11) becomes

$$E\left(k_x,k_y,k_z\right)=\frac{\hbar^2k_x^2}{2m_{11}}+\frac{\hbar^2k_y^2}{2m_{22}}+\frac{\hbar^2k_z^2}{2m_{33}}+\frac{\hbar^2k_xk_y}{m_{12}}+\frac{\hbar^2k_yk_z}{m_{23}}+\frac{\hbar^2k_zk_x}{m_{31}}. \quad (12)$$

By substituting $k_x\rightarrow -i\frac{\partial}{\partial x}$ and $k_z\rightarrow -i\frac{\partial}{\partial z}$ we now find the corresponding 2-D effective mass equation:

$$\left[-\frac{\hbar^2}{2m_{11}}\frac{\partial^2}{\partial x^2}-i\frac{\hbar^2k_y}{m_{12}}\frac{\partial}{\partial x}+\frac{\hbar^2k_y^2}{2m_{22}}+\left\{-\frac{\hbar^2}{2m_{33}}\frac{\partial^2}{\partial z^2}-i\hbar^2\left(\frac{k_y}{m_{23}}-i\frac{1}{m_{31}}\frac{\partial}{\partial x}\right)\frac{\partial}{\partial z}+W\left(x,z\right)\right\}\right]\Psi_{k_y}\left(x,z\right)=E\Psi_{k_y}\left(x,z\right)$$

(13)

Here the potential along the width direction is assumed unvarying, so that $k_y$ remains good quantum number. We now discuss the general mode-space formalism for solving this problem in general, without postulating any separability for the potential energy $W(x,z)$.

*A. The Quantum Confinement Problem:*

The confinement modes diagonalize the part of the Hamiltonian associated with the confinement potential, and serve as basis sets for evaluating the complete transport equation. From (13) we separate out the terms dealing with the quantum confinement problem at a given $x$ and find

$$\left[H_z + W(x,z)\right]\zeta_i\left(-i\frac{\partial}{\partial x}, k_y : x, z\right) = \varepsilon_i\left(-i\frac{\partial}{\partial x}, k_y : x\right)\zeta_i\left(-i\frac{\partial}{\partial x}, k_y : x, z\right) \qquad (14)$$

where the confinement Hamiltonian is

$$H_z = -\frac{\hbar^2}{2m_{33}}\frac{\partial^2}{\partial z^2} - i\hbar^2\left(\frac{k_y}{m_{23}} - i\frac{1}{m_{31}}\frac{\partial}{\partial x}\right)\frac{\partial}{\partial z} \qquad (15)$$

We now perform a canonical transformation by substituting

$$\zeta_i\left(-i\frac{\partial}{\partial x}, k_y : x, z\right) = e^{-i\left(\frac{m_{33}}{m_{23}}k_y - i\frac{m_{33}}{m_{31}}\frac{\partial}{\partial x}\right)z}\phi_i(x,z) \qquad (16)$$

in (14) and left multiplying it by $e^{i\left(\frac{m_{33}}{m_{23}}k_y - i\frac{m_{33}}{m_{31}}\frac{\partial}{\partial x}\right)z}$. The exponential term in (16) represents a unitary operator that performs a basis transformation for the wave function. The algebra is considerably simplified by employing the following operator identity:

$$e^{-B}Ae^{B} = A + [A,B] + \frac{1}{2}[[A,B],B] + ........ \qquad (17)$$

The above operation on the kinetic energy $H_z$ causes the linear terms in $\partial/\partial z$ to drop out (the exponential term being just the translation operator in $\partial/\partial z$ space), while the corresponding unitary operation on the potential $W(x,z)$ transforms it into $W\left(x+\frac{m_{33}}{m_{31}}z,z\right)$. For a given $x$, this expression implies that the quantum confinement potential needs to be sampled along the principal axis of the constant energy ellipse at fixed $k_y$. At the end of this canonical transformation, the confinement problem becomes

$$\left[-\frac{\hbar^2}{2m_{33}}\frac{\partial^2}{\partial z^2}+W\left(x+\frac{m_{33}}{m_{31}}z,z\right)\right]\phi_i\left(x,z\right)=\varepsilon_i\left(x+\frac{m_{33}}{m_{31}}z\right)\phi_i\left(x,z\right) \tag{18}$$

which we have to solve in order to obtain the orthonormal eigenvectors, $\phi_i$ s, hereafter referred to as modes. Using the inverse canonical transformation, the confinement problem can be rewritten as:

$$\left[H_z+W\left(x,z\right)\right]e^{-i\left(\frac{m_{33}}{m_{23}}k_y-i\frac{m_{33}}{m_{31}}\frac{\partial}{\partial x}\right)z}\phi_i\left(x,z\right)=\left\{\varepsilon_i\left(x\right)-\varepsilon\left(-i\frac{\partial}{\partial x},k_y\right)\right\}e^{-i\left(\frac{m_{33}}{m_{23}}k_y-i\frac{m_{33}}{m_{31}}\frac{\partial}{\partial x}\right)z}\phi_i\left(x,z\right) \tag{19}$$

where $\varepsilon_i\left(x\right)$ is the $i$-th subband energy at $x$ and

$$\varepsilon\left(-i\frac{\partial}{\partial x},k_y\right)=\frac{\hbar^2}{2}\left(-\frac{m_{33}}{m_{31}^2}\frac{\partial^2}{\partial x^2}+\frac{m_{33}}{m_{32}^2}k_y^2-2i\frac{m_{33}}{m_{31}m_{23}}k_y\frac{\partial}{\partial x}\right) \tag{20}$$

is the kinetic energy.

Eq. (19) embodies two accomplishes when compared with (14). Firstly, we have identified the transverse modes $\phi_i$ which together with the exponential prefactor selectively diagonalize the confinement part of the Hamiltonian; secondly, we see that the eigenenergy term is separated into a subband energy and a kinetic energy. In the next section, we exploit these two accomplishments.

*B. The Transport Problem:*

We now return to the original 2D effective mass equation in (13), given by

$$\left[-\frac{\hbar^2}{2m_{11}}\frac{\partial^2}{\partial x^2}-i\frac{\hbar^2k_y}{m_{12}}\frac{\partial}{\partial x}+\frac{\hbar^2k_y^2}{2m_{22}}+\overline{H_z+W\left(x,z\right)}\right]\Psi\left(-i\frac{\partial}{\partial x},k_y:x,z\right)=E\Psi\left(-i\frac{\partial}{\partial x},k_y:x,z\right) \tag{21}$$

At a given $x$, the eigenfunctions $e^{-i\left(\frac{m_{33}}{m_{23}}k_y - i\frac{m_{33}}{m_{31}}\frac{\partial}{\partial x}\right)z}\phi_i\left(x,z\right)$ diagonalize the confining Hamiltonian (19) and form a complete set. This allows us to expand the wave function $\Psi$ in (21) in this complete basis:

$$\Psi\left(-i\frac{\partial}{\partial x},k_y : x,z\right)=\sum_m e^{-i\left(\frac{m_{33}}{m_{23}}k_y - i\frac{m_{33}}{m_{31}}\frac{\partial}{\partial x}\right)z}\phi_m\left(x,z\right)\chi_m\left(x,k_y\right), \tag{22}$$

where $\chi_m\left(x,k_y\right)$ are the corresponding expansion coefficients. We substitute (22) in (21) and left multiply it by $\phi_n^*\left(x,z\right)e^{i\left(\frac{m_{33}}{m_{23}}k_y - i\frac{m_{33}}{m_{31}}\frac{\partial}{\partial x}\right)z}$, which amounts to doing a unitary transformation for the transport Hamiltonian. Using the operator identity (17) and the confinement eigenvalues from (19), equation (21) boils down to:

$$\sum_m \phi_n^*\left(x,z\right)\left[H_{trans}+\varepsilon_m\left(x+\frac{m_{33}}{m_{31}}z\right)\right]\phi_m\left(x,z\right)\chi_m\left(x,k_y\right)=E\sum_m \phi_n^*\left(x,z\right)\phi_m\left(x,z\right)\chi_m\left(x,k_y\right) \tag{23}$$

where the Hamiltonian for this transport problem is now

$$H_{trans}=-\frac{\hbar^2}{2m_1'}\frac{\partial^2}{\partial x^2}-i\frac{\hbar^2 k_y}{m_{12}'}\frac{\partial}{\partial x}+\frac{\hbar^2 k_y^2}{2m_2'}.$$

In (23) the new effective masses are obtained by regrouping terms as:

$$\frac{1}{m_1'}=\left(\frac{1}{m_{11}}-\frac{m_{33}}{m_{31}^2}\right) \tag{24a}$$

$$\frac{1}{m_2'}=\left(\frac{1}{m_{22}}-\frac{m_{33}}{m_{23}^2}\right) \tag{24b}$$

and

$$\frac{1}{m_{12}'}=\left(\frac{1}{m_{12}}-\frac{m_{33}}{m_{23}m_{31}}\right) \tag{24c}$$

Equation (23) can be simplified further using another canonical transformation:

$$\chi_m\left(x,k_y\right)=e^{-i\frac{m_1'}{m_{12}'}k_y x}\psi_m\left(x\right), \tag{25}$$

which on substituting in (23) and left multiplying with $e^{i\frac{m_1'}{m_{12}'}k_y x}$ eliminates the linear terms in $\partial/\partial$x. We find:

$$\sum_m \phi_n^*\left(x,z\right)\left[-\frac{\hbar^2}{2m_1'}\frac{\partial^2}{\partial x^2}+\frac{\hbar^2 k_y^2}{2m_2''}+\varepsilon_m\left(x+\frac{m_{33}}{m_{31}}z\right)\right]\phi_m\left(x,z\right)\psi_m\left(x\right)=E\sum_m \phi_n^*\left(x,z\right)\phi_m\left(x,z\right)\psi_m\left(x\right)$$

(26a)

where we have again made use of (17) and defined

$$\frac{1}{m_2''}=\frac{1}{m_2'}-\frac{m_1'}{m_{12}'^2} \tag{26b}$$

Equation (26a) is now integrated over $Z$. Employing the orthogonality condition

$$\int \phi_n^*\left(x,z\right)\phi_m\left(x,z\right)dz=\delta_{nm},$$

we find

$$\sum_m \int dz\left\{\phi_n^*\left(x,z\right)\left[-\frac{\hbar^2}{2m_1'}\frac{\partial^2}{\partial x^2}+\varepsilon_m\left(x+\frac{m_{33}}{m_{31}}z\right)\right]\phi_m\left(x,z\right)\right\}\psi_m\left(x\right)=\left(E-\frac{\hbar^2 k_y^2}{2m_2''}\right)\psi_n\left(x\right)$$

(27)

The potential term in (27) can be expanded by using Taylor series:

$$\begin{aligned}
\int \phi_n^*\left(x,z\right)\varepsilon_m\left(x+\frac{m_{33}}{m_{31}}z\right)\phi_m\left(x,z\right)dz &= \int \phi_n^*\left(x,z\right)\left[\varepsilon_m\left(x\right)+\frac{m_{33}}{m_{31}}z\frac{\partial \varepsilon_m\left(x\right)}{\partial x}+\ldots\right]\phi_m\left(x,z\right)dz \\
&= \varepsilon_m\left(x\right)\delta_{nm}+\frac{m_{33}}{m_{31}}\frac{\partial \varepsilon_m\left(x\right)}{\partial x}\int z\phi_n^*\left(x,z\right)\phi_m\left(x,z\right)dz+\ldots \\
&= \varepsilon_m\left(x\right)\delta_{nm}+\frac{m_{33}}{m_{31}}\frac{\partial \varepsilon_m\left(x\right)}{\partial x}\mu_{nm}+\ldots. \\
&= \varepsilon_m\left(x\right)\delta_{nm}+W_{nm}\left(x\right)
\end{aligned} \tag{28}$$

while the kinetic energy operator can be simplified using integration by parts, yielding our general coupled-mode space equation for the full 3-D transport problem:

$$\left[-\frac{\hbar^2}{2m_1'}\frac{\partial^2\psi_n(x)}{\partial x^2}+\varepsilon_n(x)\psi_n(x)\right]+\sum_m W_{nm}(x)\psi_m(x)$$

$$+\frac{\hbar^2}{2m_1'}\sum_m\left\{\psi_m(x)\int dz\left[\phi_m(x,z)\frac{\partial^2\phi_n^*(x,z)}{\partial x^2}\right]+2\int dz\frac{\partial\phi_n^*(x,z)}{\partial x}\frac{\partial\left[\phi_m(x,z)\psi_m(x)\right]}{\partial x}\right\}$$

$$=\left(E-\frac{\hbar^2k_y^2}{2m_2''}\right)\psi_n(x) \tag{29}$$

Equation (29) serves as the generalization of a similar one derived for the restricted case of Si(100) MOSFETs in [7]; however, in that treatment the summation term involving couplings $W_{nm}(x)$ among the subbands, arising due to non-alignment of device axes with ellipsoid axes, were absent. These terms represent Zener tunneling between the subbands, and ignoring the higher order corrections, they are proportional to the intersubband dipole $\mu_{nm}$ and the local field generated by the variation in the subband eigenvalues along the transport direction. Contributions to $W_{nm}(x)$ arise both from cross terms in the potential energy, as in a channel with varying cross-section, as well as from cross-terms in the kinetic energy due to the non-aligned device and ellipsoidal axes. While the latter terms never arise for Si (100) devices, the effect of the former cross terms is included in the coupled mode space approach of [13] to analyze effects of channel access geometry in nanoscale Si n-MOSFETs.

*C. Simplifications- Uniform Cross-section, Separable Potential and Ultra Thin Body*

In an UTB SOI MOSFET with uniform channel thickness (Fig. 2), the electrostatic potential is separable along the confinement and transport (channel) directions:

$$W(x,z)=U(z)+V(x) \tag{30a}$$

The channel potential $V(x)$ simply shifts the confining potential $U(z)$, but does not alter the shape of the modes along the transport direction. This implies

$$\frac{\partial\phi_j(x,z)}{\partial x}=\frac{\partial^2\phi_j(x,z)}{\partial x^2}=0\,. \tag{30b}$$

The $x$ dependence of $\phi_j(x,z)$ is eliminated since it selectively diagonalizes the confinement Hamiltonian which depends only on $U(z)$ and is therefore $x$-independent. In this special case (29) simplifies considerably and becomes

$$\left[-\frac{\hbar^2}{2m_1'}\frac{\partial^2\psi_n(x)}{\partial x^2}+\varepsilon_n(x)\psi_n(x)\right]+\sum_m W_{nm}(x)\psi_m(x)=\left(E-\frac{\hbar^2k_y^2}{2m_2''}\right)\psi_n(x)\cdot \tag{31a}$$

Additionally, (30a) allows us to write

$$\varepsilon_n\left(x+\frac{m_{33}}{m_{31}}z\right)=\varepsilon_n+V\left(x+\frac{m_{33}}{m_{31}}z\right) \tag{31b}$$

where $\varepsilon_n$ is the $n$-th subband energy arising from diagonalizing the $x$-independent confinement potential, and is therefore is unvarying along $x$. The only role of the channel potential $V(x)$ is to shift the bottoms of the subbands.

$$\left[-\frac{\hbar^2}{2m_1'}\frac{\partial^2}{\partial x^2}+V(x)\right]\psi_n(x)+\sum_{m\neq n}V_{nm}(x)\psi_m(x)=\left(E-\varepsilon_n-\frac{\hbar^2k_y^2}{2m_2''}\right)\psi_n(x) \tag{32}$$

In (32) the coupling between modes is still present and represents Zener tunneling between subbands due to cross-terms in the kinetic energy, arising from the arbitrary orientation of the ellipsoids. This tunneling is negligible if

$$\left|V_{nm}(x)\right|<<\left|\varepsilon_n(x)-\varepsilon_m(x)\right|$$

which amounts to

$$\left|\frac{m_{33}}{m_{31}}\frac{\partial V(x)}{\partial x}\mu_{nm}\right|<<\left|\varepsilon_n-\varepsilon_m\right| \tag{33}$$

this inequality can be restated as follows: **if at any point along the channel, the channel directed potential drop between the confining planes and sampled along the ellipsoidal principal axis direction is much smaller than the corresponding intersubband separation, the coupling between modes can be safely ignored**. Indeed, the coupling arose precisely because the channel potential varies along the ellipsoidal direction, which determines the confinement potential generating the transverse subbands. For Si(100) the coupling does not exist since there is no drop in the channel directed potential along the direction of the conduction band ellipsoids.

A simple estimate tells us the conditions under which this Zener tunneling is negligible. The smallest intersubband separation is given roughly by $(3^2 - 1^2)\frac{\hbar^2 \pi^2}{2m_{33} t^2}$, $t$ being the body thickness. The channel potential drop along the ellipsoid between the confinement planes depends on the local field. For a linear potential profile, this drop is given by $\frac{Vt}{L}$, where $V$ is the applied bias and $L$ is the channel length. Near the top of the barrier the field is smaller and the drop is given roughly by $\frac{Vt^2}{2L^2}$. For a 10 nm channel length with a 0.6 volt applied bias and for $m_{33}$=0.1, the drop in channel potential along the confinement direction is negligible compared to the subband separation provided the body thickness is smaller than about 5 nm (for the average potential, and 10 nm for the top of the barrier potential).

In the absence of intersubband Zener coupling, the corresponding decoupled mode-space equation finally becomes

$$-\frac{\hbar^2}{2m_1'}\frac{\partial^2 \psi_n(x)}{\partial x^2} + V(x)\psi_n(x) = \left(E - \varepsilon_n - \frac{\hbar^2 k_y^2}{2m_2''}\right)\psi_n(x). \tag{34}$$

In this section we have demonstrated an exact mathematical transformation that considerably simplifies the quantum mechanical treatment of electronic transport in a nanoscale n-MOSFET with arbitrarily oriented transport, width and thickness directions. The conditions under which this simplified equation is valid are (a) an unvarying device cross-section that allows us to separate the confinement and transport potentials, and (b) a transport potential that varies slowly enough that there is no Zener tunneling between transverse subbands. The simplified mapping operation is performed by two unitary operations in (16) and (25) that map each conduction band ellipsoid into an equivalent regular ellipsoid whose principal axes are oriented along the device coordinate axes, $X$, $Y$ and $Z$ with corresponding effective masses, $m_1'$, $m_2''$ and $m_{33}$, respectively. For separable potentials with coincident device and ellipsoidal axes (i.e., no cross terms in the kinetic energy), the normal modes $\phi_m(z)$ selectively diagonalize the confinement Hamiltonian $-\partial^2/\partial z^2 + U(z)$. Non-coincident device and ellipsoidal directions introduces additional cross terms in the Hamiltonian which are diagonalized by $e^{-i\left(\frac{m_{33}}{m_{23}}k_y - i\frac{m_{33}}{m_{31}}\frac{\partial}{\partial x}\right)z}\phi_m(x,z)$, while in the general case of an arbitrary potential $W(x,z)$, the transverse modes $\phi_m(x,z)$ depend on both *x* and *z*. In the next two sections, we present several applications of the transformation developed here.

## V. APPLICATION TO UTB SI AND GE MOSFETS

We now demonstrate the usefulness of our generalized treatment by applying it to several technologically important materials and wafer orientations. Until recently nearly all quantum simulations of MOSFETs were performed for silicon $(100)$ wafers with transport along one of the other two $[100]$ directions. We will discuss several nontrivial cases here — effective masses for ultra-thin body (UTB) silicon and germanium MOSFETs fabricated on $(100)$, $(111)$ and $(110)$ wafers.

### *A. Transformation matrices for the Device coordinate system (DCS)*

First we will evaluate the transformation matrix $\Re_{C\leftarrow D}$ for various wafer orientations.

#### *A1.* $(100)$ *wafers:*

This is the commonest wafer orientation used for the fabrication and simulation of nanoscale silicon MOSFETs. In these devices the inversion layer electrons are confined along the $[001]$ direction which is the $Z$ axis for the DCS. The transport and width directions ($X$ and $Y$ axes) are along $[100]$ and $[010]$, respectively, so that the basis vectors are $\hat{k}_1 = (1,0,0)$, $\hat{k}_2 = (0,1,0)$ and $\hat{k}_3 = (0,0,1)$. Since the columns of the transformation matrix $\Re_{C\leftarrow D}$ are components of $\hat{k}_1$, $\hat{k}_2$ and $\hat{k}_3$, the matrix itself becomes an identity matrix:

$$\Re^{(001)}_{C\leftarrow D}=\begin{bmatrix}1 & 0 & 0\\ 0 & 1 & 0\\ 0 & 0 & 1\end{bmatrix} \tag{36}$$

*A2.* $(111)$ *wafers:*

For MOSFETs fabricated on $(111)$ wafers, the gate electric field confines the inversion layer carriers along $[111]$ crystallographic orientation. We choose the transport direction along $[\bar{2}11]$ and the width direction along $[0\bar{1}1]$. These crystallographic orientations represent the $Z$, $X$ and $Y$ axes for the device. The basis vectors for DCS are $\hat{k}_1=\left(-\frac{2}{\sqrt{6}},\frac{1}{\sqrt{6}},\frac{1}{\sqrt{6}}\right)$, $\hat{k}_2=\left(0,-\frac{1}{\sqrt{2}},\frac{1}{\sqrt{2}}\right)$ and $\hat{k}_3=\left(\frac{1}{\sqrt{3}},\frac{1}{\sqrt{3}},\frac{1}{\sqrt{3}}\right)$, respectively. The rotation matrix $\Re_{C\leftarrow D}$ becomes

$$\Re^{(111)}_{C\leftarrow D}=\begin{bmatrix}-2/\sqrt{6} & 0 & 1/\sqrt{3}\\ 1/\sqrt{6} & -1/\sqrt{2} & 1/\sqrt{3}\\ 1/\sqrt{6} & 1/\sqrt{2} & 1/\sqrt{3}\end{bmatrix}, \tag{37}$$

*A3.* $(110)$ *wafers:*

In this case the inversion layer electrons are confined along $[110]$ crystallographic orientation, which is the $Z$ axis. We choose the transport ($X$) direction along $[001]$ and the width ($Y$) direction along $[1\bar{1}0]$, so that the corresponding unit vectors are $\hat{k}_1=(0,0,1)$, $\hat{k}_2=\left(\frac{1}{\sqrt{2}},\frac{-1}{\sqrt{2}},0\right)$ and $\hat{k}_3=\left(\frac{1}{\sqrt{2}},\frac{1}{\sqrt{2}},0\right)$, and the rotation matrix $\Re_{C\leftarrow D}$:

$$\Re^{(110)}_{C\leftarrow D}=\begin{bmatrix}0 & 1/\sqrt{2} & 1/\sqrt{2}\\ 0 & -1/\sqrt{2} & 1/\sqrt{2}\\ 1 & 0 & 0\end{bmatrix}. \tag{38}$$

*B. Transformation matrices for the Ellipsoidal coordinate system (ECS)*

There are two types of valleys in indirect bandgap semiconductors — the sixfold degenerate Δ valleys and the eightfold degenerate Λ valleys. These are classified according to the orientation of the major axes of the constant energy ellipsoids, along the $\langle 100\rangle$ or $\langle 111\rangle$ directions. In bulk silicon the Δ valleys are energetically lower than the Λ valleys; consequently, the conduction band electrons populate the Δ valleys, and the Λ valleys can be ignored for transport simulations. The opposite is true for bulk germanium, where the Λ valleys are energetically lower than their Δ counterparts, and therefore, the states near the conduction band edge are of the former type. Interesting phenomena can be observed when quantum confinement is present, since there in addition to the energy of the bulk band edge, inversion layer thickness and confinement direction effective mass determine which valley forms the energetically lowest subband. In this subsection, we will evaluate the transformation matrix , $\Re_{E\leftarrow C}$, for these valleys.

*B1.* Δ- *valleys:*

Fig. 3 shows the three doubly degenerate constant energy Δ valley conduction band ellipsoids. The basis vectors for the ellipsoid coordinate system are different for each ellipsoid, with $\hat{k}_{\parallel}$ along the major axis, and $\hat{k}_{\perp 1}$ and $\hat{k}_{\perp 2}$ along two orthogonal minor axes. For ellipsoid 1, for example, $\hat{k}_{\parallel}=(1,0,0)$, $\hat{k}_{\perp 1}=(0,1,0)$ and $\hat{k}_{\perp 2}=(0,0,1)$. For each ellipsoid there is an unique transformation matrix $\Re_{E\leftarrow C}$, the rows of which are the components of $\hat{k}_{\parallel}$, $\hat{k}_{\perp 1}$and $\hat{k}_{\perp 2}$. For ellipsoid 1, $\Re_{E\leftarrow C}$ becomes

$$\Re^{\Delta_1}_{E\leftarrow C}=\begin{bmatrix}1 & 0 & 0\\ 0 & 1 & 0\\ 0 & 0 & 1\end{bmatrix}. \tag{39a}$$

Similar expressions can be obtained for ellipsoids 2 and 3:

$$\Re^{\Delta_2}_{E\leftarrow C}=\begin{bmatrix}0 & 1 & 0\\ 0 & 0 & 1\\ 1 & 0 & 0\end{bmatrix}, \tag{39b}$$

and

$$\Re^{\Delta_3}_{E\leftarrow C} = \begin{bmatrix} 0 & 0 & 1 \\ 1 & 0 & 0 \\ 0 & 1 & 0 \end{bmatrix}. \tag{39c}$$

*B2. Λ- valleys*

Figure 4a shows the eight half Λ-valley ellipsoids with centers at equivalent L points at the surface of first Brillouin zone. Since the diagonally opposite L points are one reciprocal lattice vector apart, they can be combined into four full ellipsoids, as shown in Fig. 4b. The major axes of ellipsoids 1-4 are along $[111]$, $[11\bar{1}]$, $[\bar{1}11]$ and $[\bar{1}1\bar{1}]$, respectively. The rotation matrix $\Re^{\Lambda_1}_{E\leftarrow C}$ for ellipsoid 1 can be written from the components of the basis vectors $\hat{k}_{\parallel}$, $\hat{k}_{\perp 1}$ and $\hat{k}_{\perp 2}$ in the CCS, and is,

$$\Re^{\Lambda_1}_{E\leftarrow C} = \begin{bmatrix} \frac{1}{\sqrt{3}} & \frac{1}{\sqrt{3}} & \frac{1}{\sqrt{3}} \\ -\frac{1}{\sqrt{2}} & \frac{1}{\sqrt{2}} & 0 \\ -\frac{1}{\sqrt{6}} & -\frac{1}{\sqrt{6}} & \frac{2}{\sqrt{6}} \end{bmatrix}. \tag{40a}$$

Similar matrices for ellipsoids 2-4 can be readily calculated, and are

$$\Re^{\Lambda_2}_{E\leftarrow C} = \begin{bmatrix} \frac{1}{\sqrt{3}} & \frac{1}{\sqrt{3}} & -\frac{1}{\sqrt{3}} \\ -\frac{1}{\sqrt{2}} & \frac{1}{\sqrt{2}} & 0 \\ \frac{1}{\sqrt{6}} & \frac{1}{\sqrt{6}} & \frac{2}{\sqrt{6}} \end{bmatrix}, \tag{40b}$$

$$\Re^{\Lambda_3}_{E\leftarrow C} = \begin{bmatrix} -\frac{1}{\sqrt{3}} & \frac{1}{\sqrt{3}} & \frac{1}{\sqrt{3}} \\ \frac{1}{\sqrt{2}} & \frac{1}{\sqrt{2}} & 0 \\ -\frac{1}{\sqrt{6}} & \frac{1}{\sqrt{6}} & -\frac{2}{\sqrt{6}} \end{bmatrix}, \tag{40c}$$

and

$$\Re^{\Lambda_4}_{E\leftarrow C} = \begin{bmatrix} -\frac{1}{\sqrt{3}} & \frac{1}{\sqrt{3}} & -\frac{1}{\sqrt{3}} \\ \frac{1}{\sqrt{2}} & \frac{1}{\sqrt{2}} & 0 \\ \frac{1}{\sqrt{6}} & -\frac{1}{\sqrt{6}} & -\frac{2}{\sqrt{6}} \end{bmatrix}. \tag{40d}$$

*C. Evaluating effective masses*

Using the results of the previous two subsections, we can now calculate the effective masses for both Δ and Λ valleys in the conduction band, and for different wafer orientations. In Table I, the results are given in terms of bulk $m_l$ and $m_t$. The following steps were used to obtain the results:

**Step 1:** For the given wafer orientation, choose the appropriate $\Re_{C\leftarrow D}$ from (36)-(38).

**Step 2:** For the given valley type (Δ or Λ) and for each of the conduction band ellipsoids, choose the appropriate $\Re_{E\leftarrow C}$ from (39)-(40).

**Step 3:** From (7), evaluate $\Re_{E\leftarrow D}$, and then using (10) find the effective mass tensor, $\left[M_D^{-1}\right]$, in the device coordinate system.

**Step 4:** The confinement effective mass, $m_Z\left(=m_{33}\right)$, is directly obtained from $\left[M_D^{-1}\right]$. The transport effective mass, $m_X\left(=m'_1\right)$, is calculated from (24a). Finally, the effective mass along width direction, $m_Y\left(=m''_2\right)$, is calculated from (26b).

Table II shows the $m_l$ and $m_t$ values for the Δ and Λ valleys of bulk silicon and bulk germanium. The effective masses of the lowest valleys for each material, i.e. Δ—valleys for silicon and Λ—valleys for germanium were obtained from cyclotron experiments [24, 25, 29, 30], while those for the upper valleys are obtained from empirical pseudopotential calculations [31, 32]. Using the effective masses in Table II and the expressions for $m_X$, $m_Y$ and $m_Z$ in Table I, the effective masses for the Δ—valleys and Λ—valleys in silicon and germanium can be calculated for the corresponding wafer orientation.

## V. RESULTS AND DISCUSSION

In this paper we introduced a generalized effective mass equation framework for the quantum mechanical simulation of cubic semiconductor n-MOSFETs. It is well known that when one or more principal axes of the conduction band constant energy ellipsoid do not coincide with the device coordinate axes, $X$, $Y$, and $Z$, the solution of effective mass equation is nontrivial. The treatment simplifies if the electrostatic potential is separable, valid if the cross-section is unvarying along the transport direction. Further simplifications occur for thin body MOSFETs, where the cross-terms in the kinetic energy arising due to the arbitrarily oriented conduction band ellipsoids do not couple the various subbands. The unitary operation in (16) decouples the energy along the confinement direction ( $Z$ axis) from the energy associated with the carrier's motion in the transport plane. A physical picture of the result of this operation is schematically presented in Fig. 5, where we see that this transforms the arbitrarily oriented ellipsoid (top) in such a way, that the resultant ellipsoid (middle) becomes symmetric across the $X-Y$ plane and, and, therefore, one principal axis of the transformed ellipsoid becomes aligned with the $Z$ axis. Consequently, in (19) we see that the $X-Y$ plane energy is decoupled from quantum confinement problem.

By substituting $-i\frac{\partial}{\partial x} \rightarrow k_x$ in (20), we find the constant energy contours for the electrons in the $X-Y$ plane as ellipses. In general, their principal axes are not aligned with $X$ and $Y$, and the Hamiltonian in (23) remains complicated due to the presence of first derivative. The second unitary operation in (25) transforms these ellipses in such a way that the transformed constant energy elliptical contours have their principal axis aligned along the transport direction. This is also schematically presented in Fig. 5, where we see that the bottom ellipsoid has its principal axes along the device axes. In summary, since the top and the bottom ellipsoids in Fig. 5 are exactly equivalent, the well defined effective masses, $m_X$, $m_Y$ and $m_Z$, determined from bottom ellipsoid describes the effective masses of the original ellipsoid. It can be observed from Table I, that for each row $m_X m_Y m_Z = m_l m_t^2$, which ensures that the volume of the transformed ellipsoid is same as that of original one, and therefore, the density-of-states is conserved. Additionally, these unitary operations change only the phase velocity of the states and thus the group velocity of the carriers, determined from the gradient of the $E\left(\vec{k}\right)$, is also conserved. Conservation of density-of-states and carrier group velocity ensures that the results obtained by performing quantum simulation of a MOSFET using, $m_X$, $m_Y$ and $m_Z$, gives the same result as from treating the non-diagonal effective mass tensor, $\left[M_D^{-1}\right]$. The density-of-states effective mass per valley can be readily obtained from Table. I, using the expression, $m_d = \sqrt{m_X m_Y}$.

Several decoupled-mode-space application of the theory developed above is presented in Figs. 6-8. In Fig. 6a-b, the electronic subband energies in a 3nm thick germanium n-MOSFET are shown. Two different wafer orientations, $(100)$ and $(111)$, are considered. The subbands for the $(100)$ wafer orientation are fourfold degenerate. For the $(111)$ wafer, since there is two different confinement effective masses, $m_Z$, we observe two ladders of subbands—unprimed and primed. The higher value of $m_Z$ corresponds to singly degenerate lower (unprimed) ladder while the lower value of $m_Z$ corresponds to the threefold degenerate upper (primed) ladder.

In Fig. 7, the on-state ballistic output characteristics of $(100)$ and $(111)$ ultra-thin-body double gate germanium n-MOSFETs are compared. The quasi-2D analytical model for nanoscale MOSFETs, described in [15], is used. The effective masses were obtained from Table I. It can be observed that due to low density-of-states effective mass in the $(111)$ n-MOSFETs, the on-current is degraded compared to their $(100)$ counterparts. Finally, in Fig. 8, the local density of states plots in an ultra-thin-body double gate germanium $(100)$ n-MOSFET are shown. The 2D quantum simulator, nanoMOS 2.5, was modified with the appropriate effective masses and NEGF formalism was used in this regard. A detailed study of germanium n-MOSFETs is published elsewhere [33].

## VI. SUMMARY AND CONCLUSION

The simple technique for mapping arbitrarily oriented conduction band constant energy ellipsoids into regular ellipsoids, where the principal axes are aligned along the device axes, allows us to perform quantum transport simulation in ultra-thin body n-MOSFETs with any channel materials and arbitrary wafer orientations. The effective masses presented in Table I can be readily used in any quantum mechanical simulator to determine I-V and C-V characteristics.

## ACKNOWLEDGEMENT

The authors would like to thank Prof. Supriyo Datta, Steve Laux, Gerhard Klimeck, Ashraf Alam, Ramesh Venugopal, and Low Aik Seng Tony (National University of Singapore) for stimulating discussions and thoughtful suggestions. This work was supported by the Semiconductor Research Corporation (SRC) under contract no. 1042.001.

# LIST OF FIGURES

according to their confinement effective mass $m_Z$. Singly degenerate unprimed ladder and triply degenerate primed ladder.

**Figure 7** The on-state ($V_{GS} = V_{DD}$ = 0.4V) ballistic output characteristics of $(100)$ and $(111)$ UTB DG Ge n-MOSFETs are compared using the quasi-2D analytical model described in [15]. The low density-of-states effective mass in the $(111)$ device is responsible for its on-current degradation.

**Figure 8** The local density of states plots in an UTB DG Ge $(100)$ n-MOSFET from the 2D quantum simulator nanoMOS 2.5. The simulator was modified according to the effective masses and valley degeneracies presented in Table I. Body thickness 2.5nm and gate length 10nm. The interference of contact injected flux and reflected flux is clearly visible. (a) Low source-to-channel barrier in on-state and (b) high barrier in the off-state.

# List of Tables

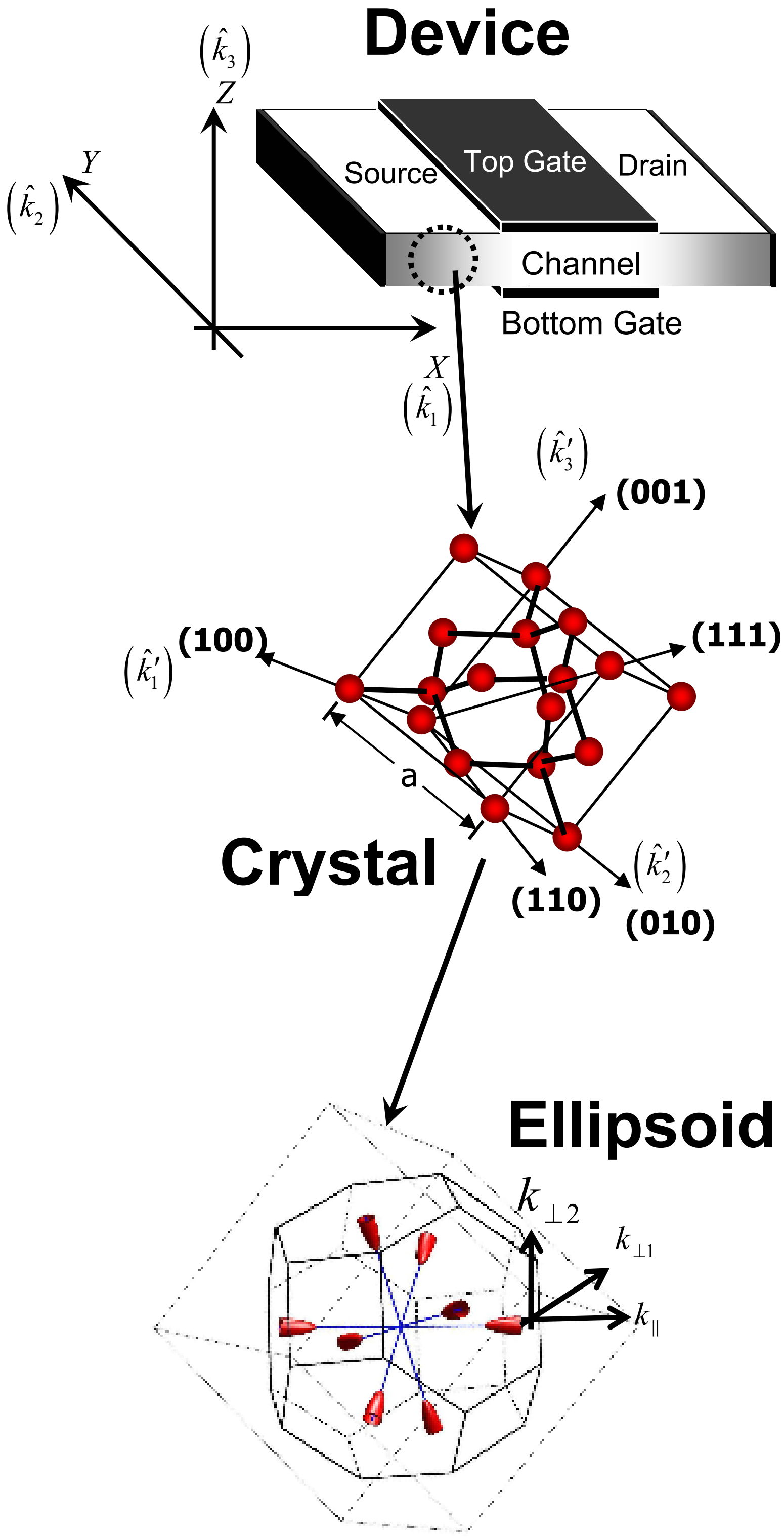

**Figure 1** Three orthogonal coordinate systems: Device coordinate system (DCS), crystal coordinate system (CCS), and ellipsoidal coordinate system (ECS).

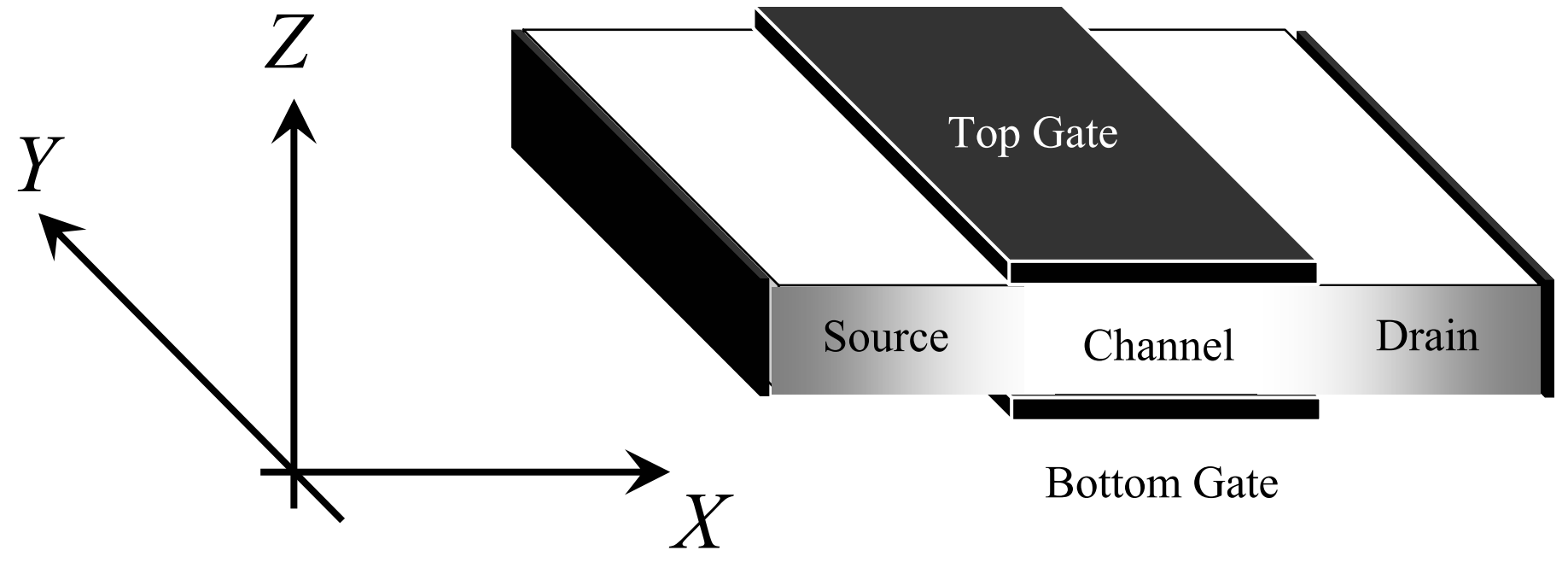

**Figure 2** The ultra-thin-body, double-gate device structure. The device coordinate system consists of orthogonal axes *X*, *Y* and *Z* along transport, width and thickness directions, respectively. The wavevectors along the thickness, $k_3$, and along the transport, $k_1$, are treated quantum mechanically while in the width direction, *Y*, plane waves were assumed.

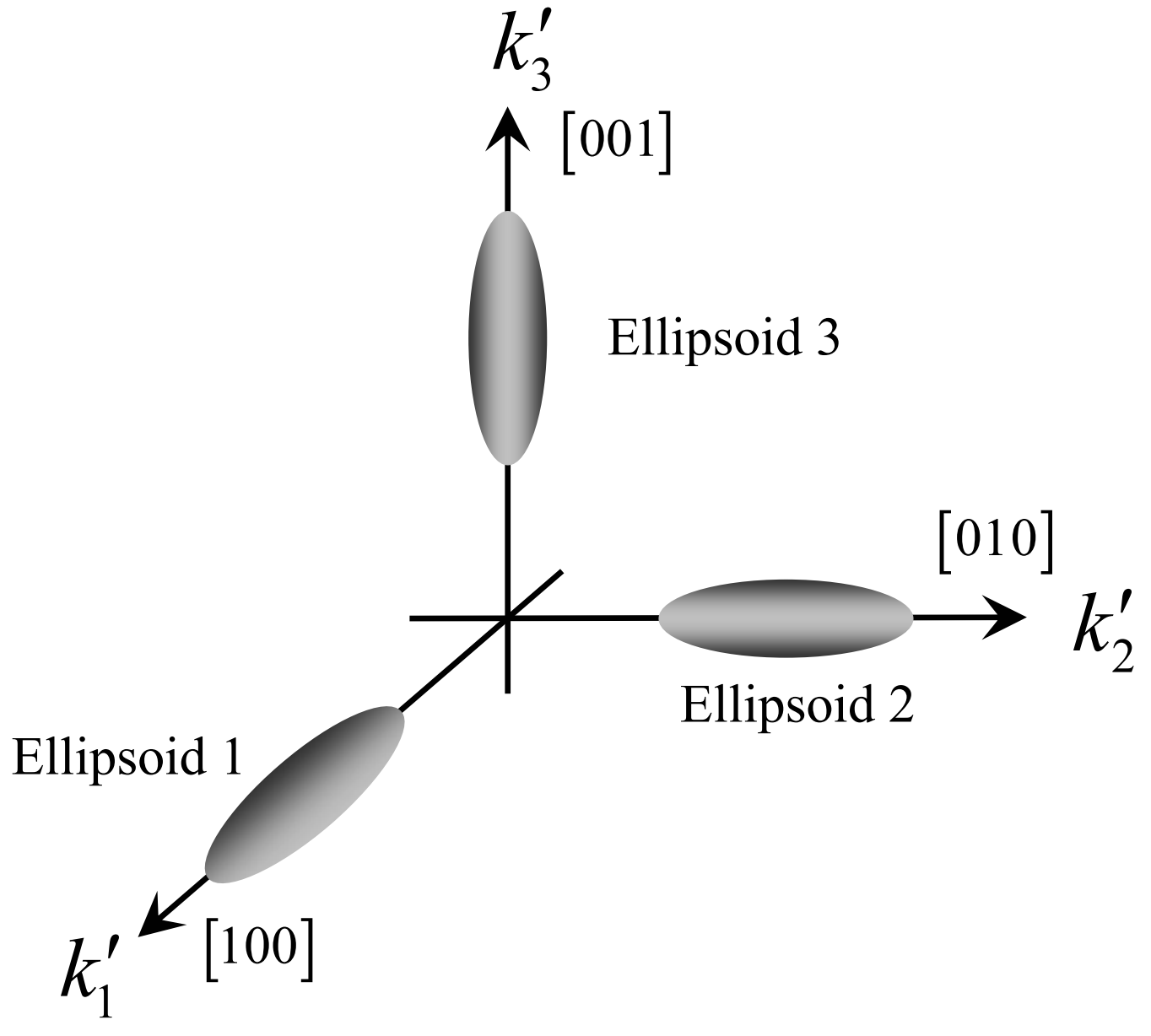

**Figure 3** Conduction band constant energy ellipsoids in along Δ. Each of the three ellipsoids shown is doubly degenerate. In silicon, such valleys form the conduction band minima.

(a)

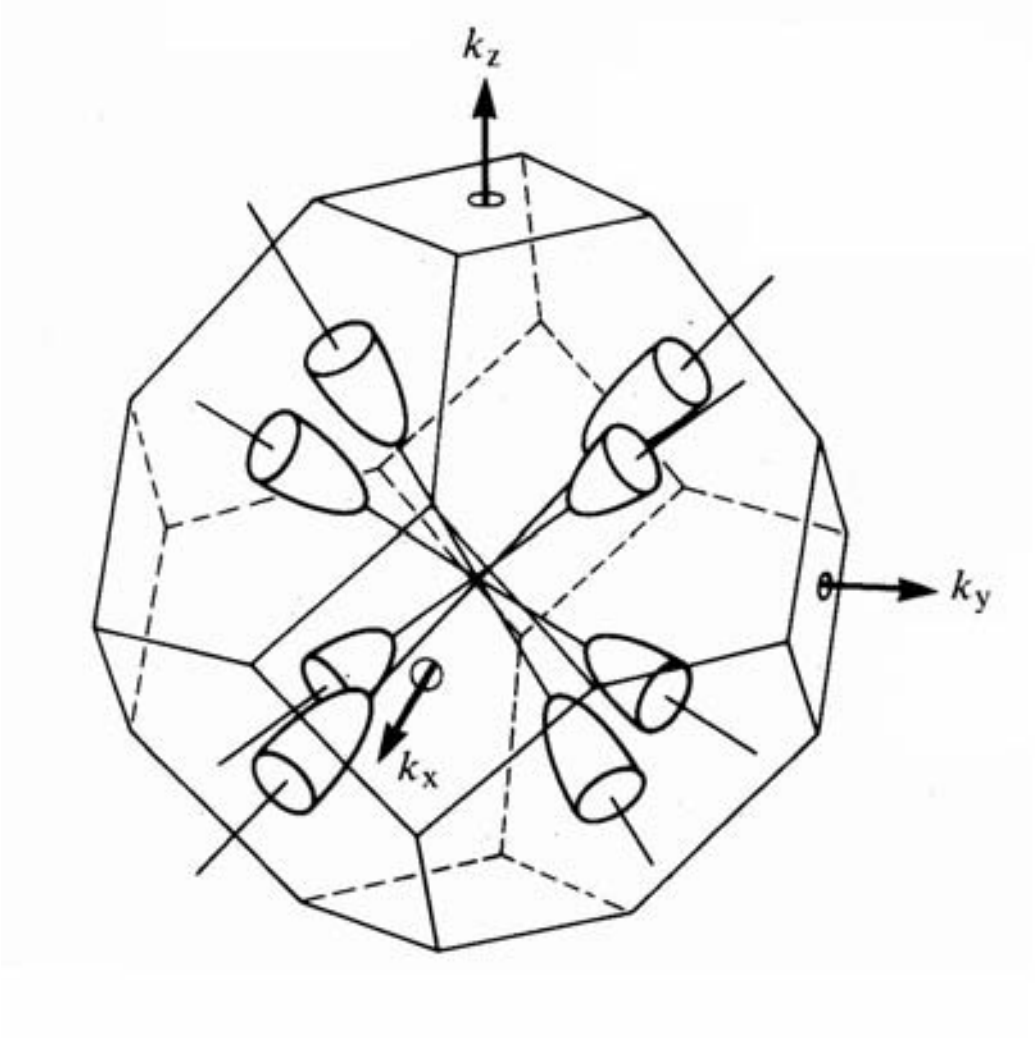

(b)

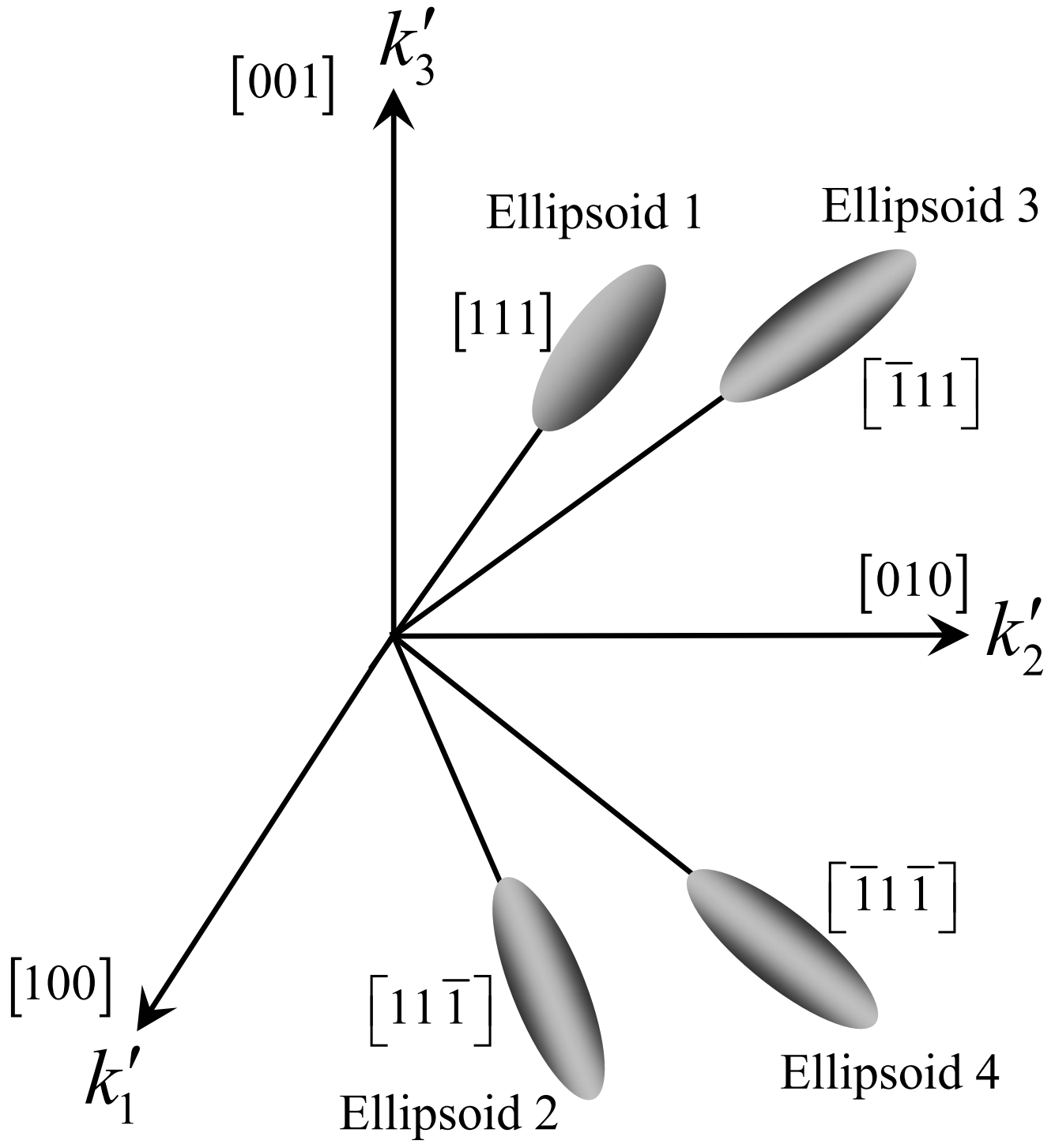

**Figure 4** Conduction band constant energy ellipsoids around the *L* points in the first BZ. (a) The major axis of the eight half ellipsoids are along Λ. (b) Since the centers of the diagonally opposite half ellipsoids are one wavevector apart, they can be combined into four equivalent full ellipsoids. In bulk Ge they form the conduction band edge.

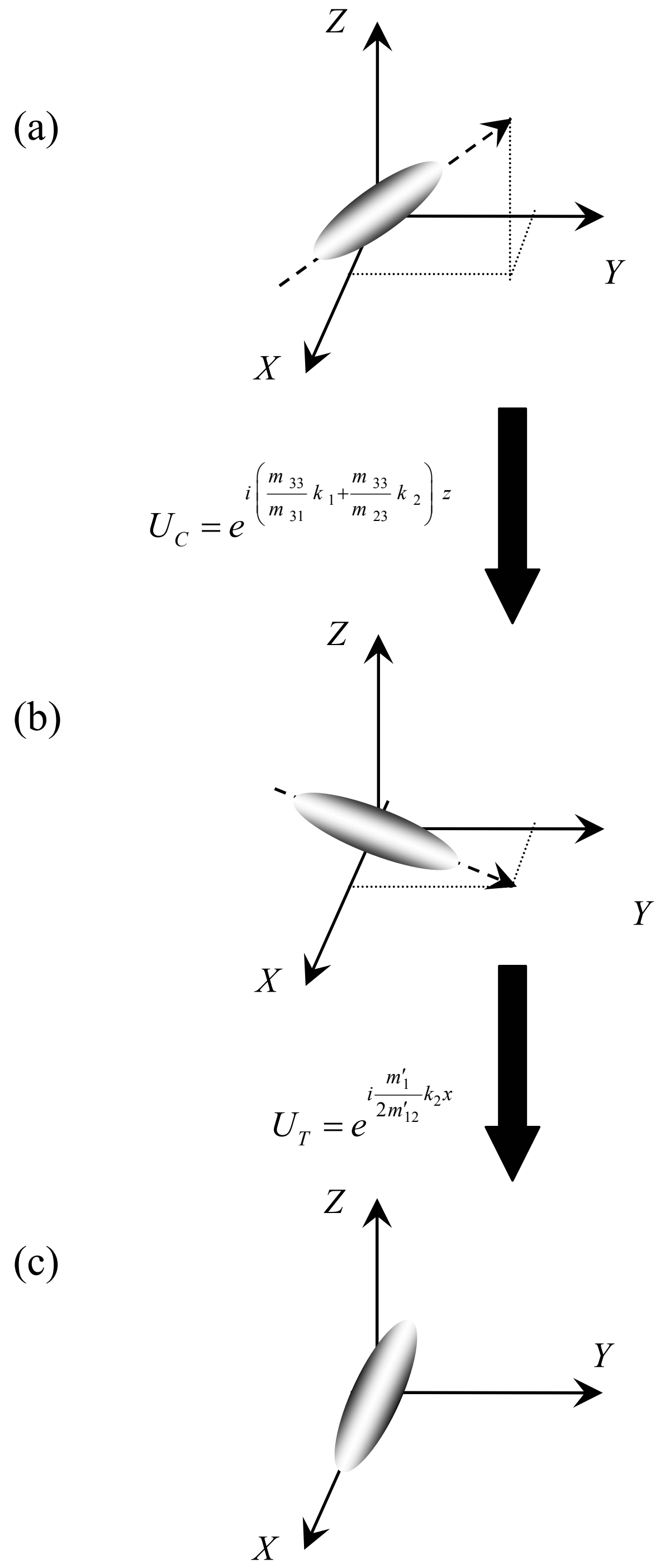

**Fig. 5** The effects of unitary transformations of eqs. (16) and (25). The first operator transforms the arbitrarily oriented CB ellipsoid in (a) into an equivalent one in (b) which is symmetric across $k_x$-$k_y$ plane. The second unitary operation transforms it into the regular ellipsoid of (c) with its principal axes along *X*, *Y* and *Z*. The density-of-states effective mass and group velocity of each *k*-state is conserved.

(a)

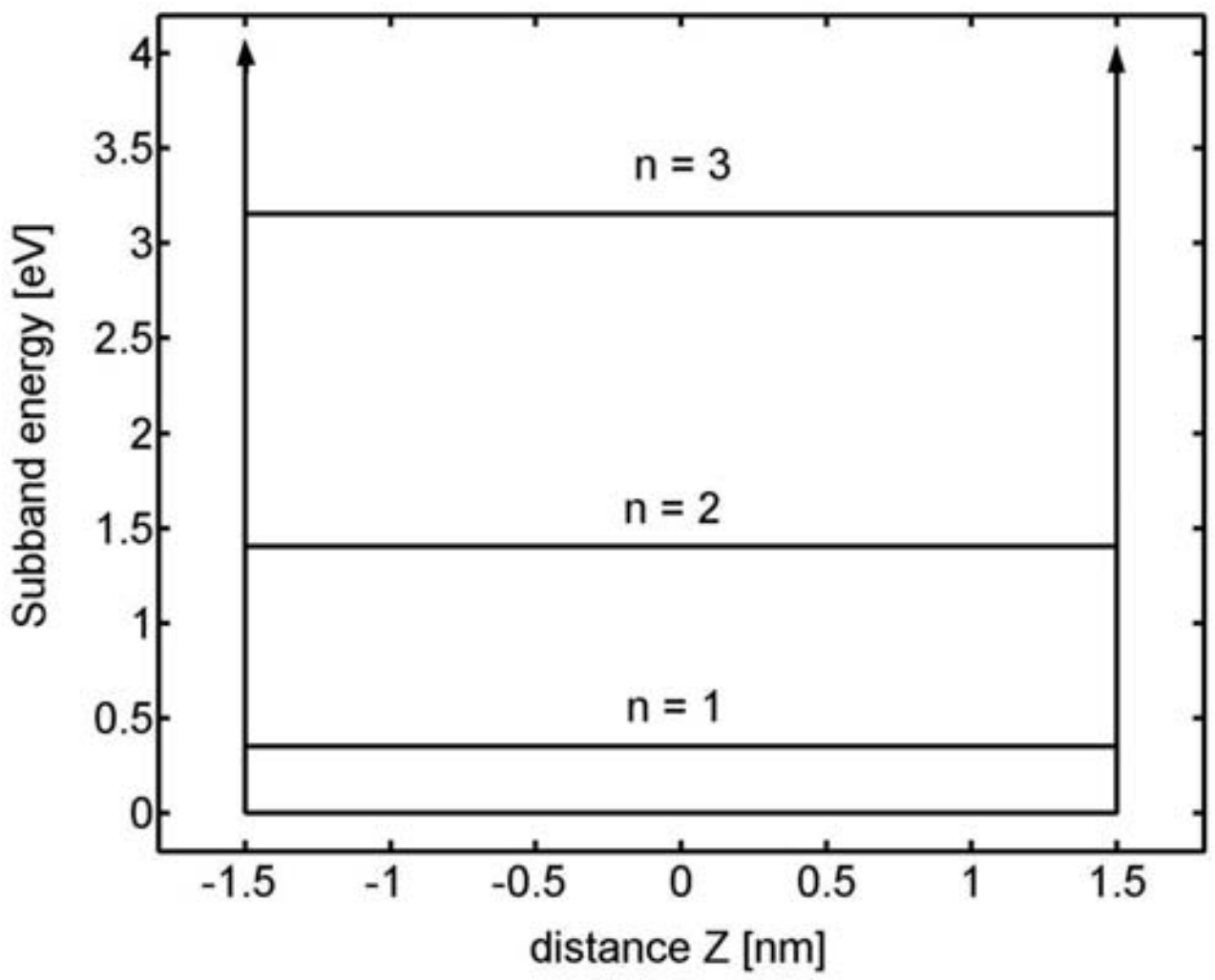

(b)

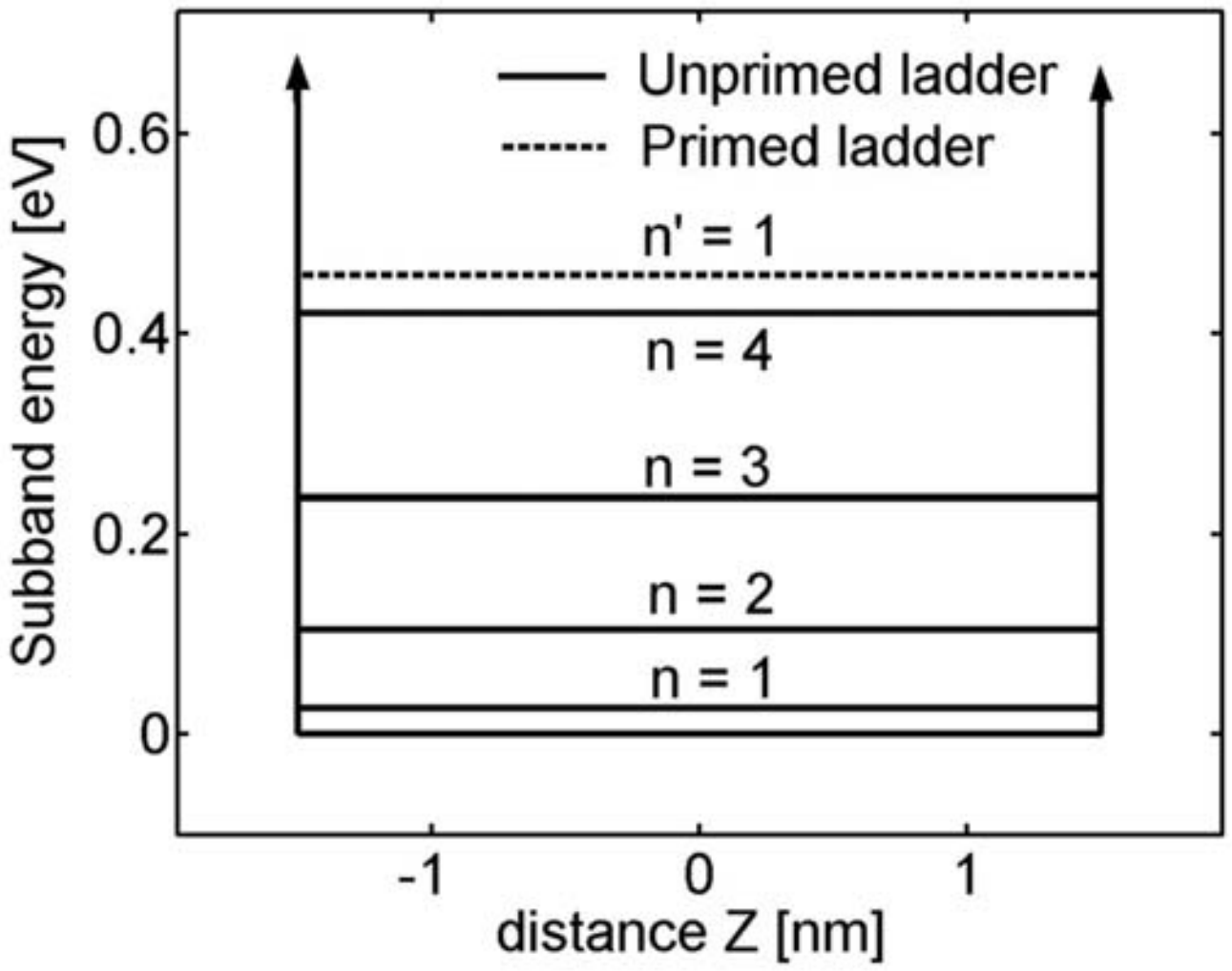

**Figure 6** The subband energies for UTB DG Ge MOSFETs for two different wafer orientations. (a) For Ge (100) wafer there is only one ladder of subbands. Each subband is fourfold degenerate. (b) For Ge (111) wafer, the fourfold degenerate *L*-type conduction band minima is separated in two groups according to their confinement effective mass $m_Z$: Singly degenerate unprimed ladder and triply degenerate primed ladder.

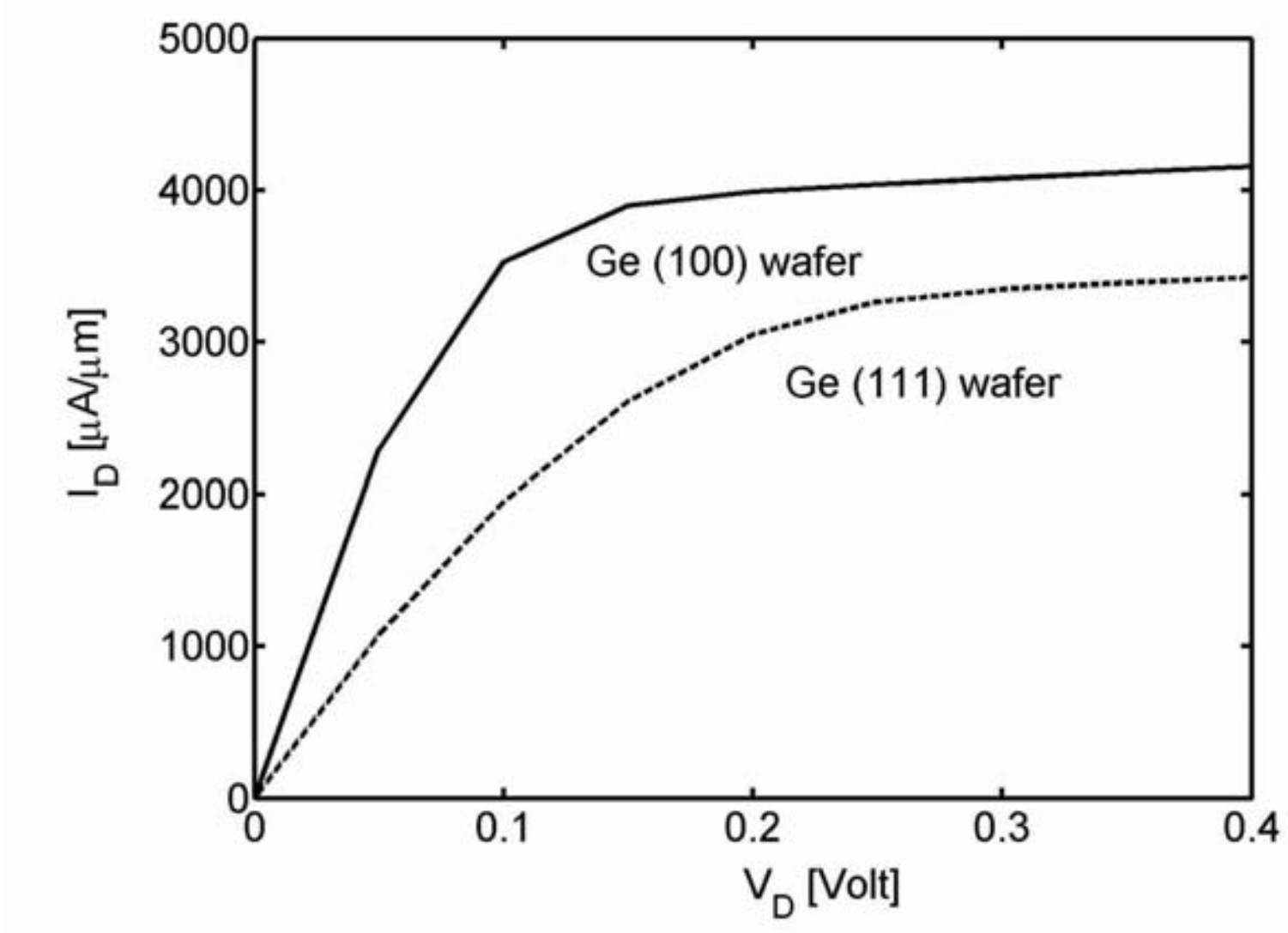

**Figure 7** The on-state ($V_{GS}=V_{DD}$ = 0.4V) ballistic output characteristics of $(100)$ and $(111)$ UTB DG Ge n-MOSFETs are compared using the quasi-2D analytical model described in [15]. The low density-of-states effective mass in the $(111)$ device is responsible for its on-current degradation.

(a)

(b)

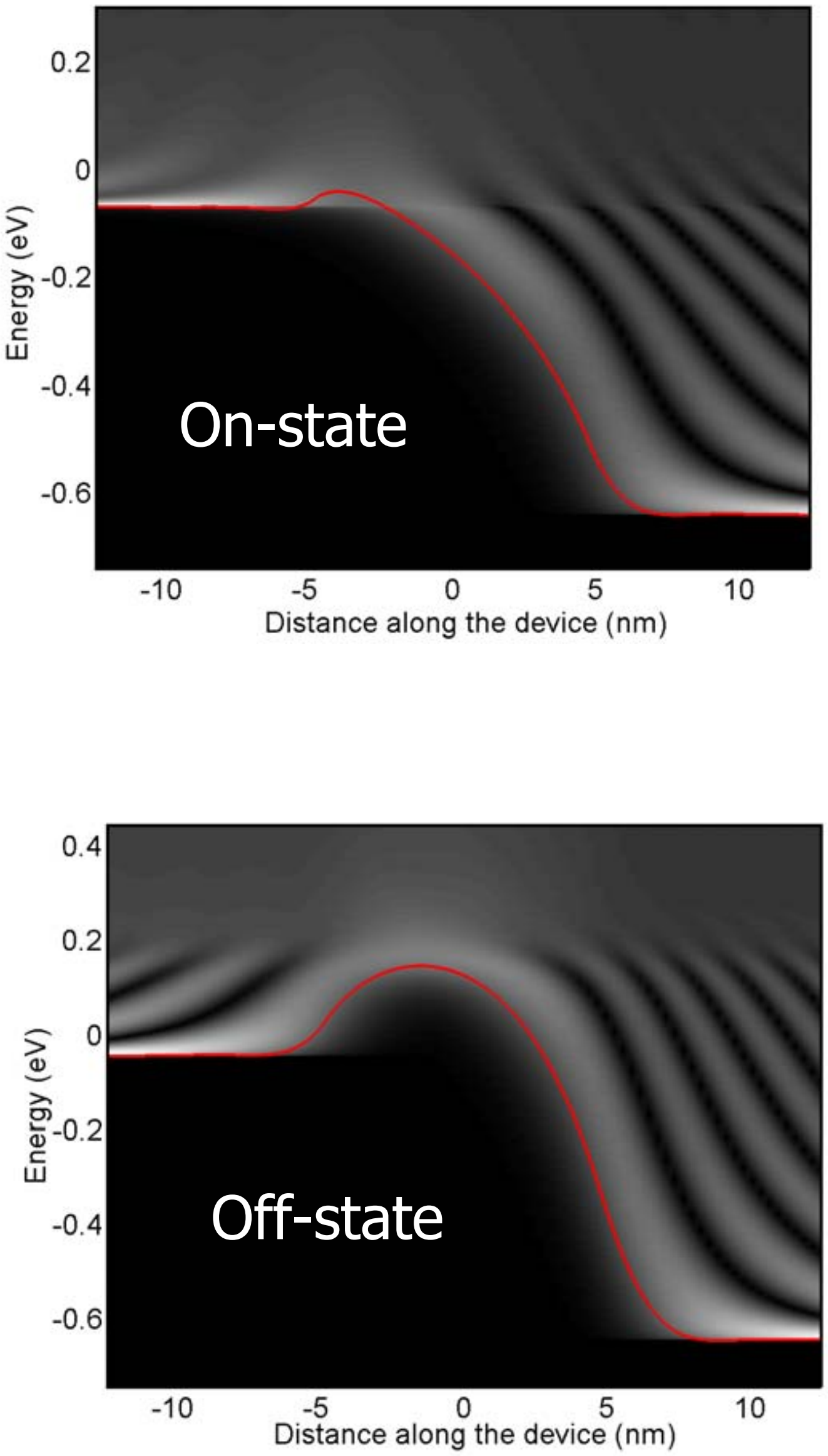

**Figure 8** The local density of states plots in an UTB DG Ge $(100)$ n-MOSFET from the 2D quantum simulator nanoMOS 2.5. The simulator was modified according to the effective masses and valley degeneracies presented in Table I. Body thickness 2.5nm and gate length 10nm. The interference of contact injected flux and reflected flux is clearly visible. (a) Low source-to-channel barrier in on-state and (b) high barrier in the off-state.

| **(Wafer)/[transport]/[width]** | Valley | $m_X$ | $m_Y$ | $m_Z$ | Degeneracy |
|---|---|---|---|---|---|
| $(001)/[100]/[010]$ | $\Delta$ | $m_t$ | $m_t$ | $m_l$ | 2 |
| | $\Delta$ | $m_l$ | $m_t$ | $m_t$ | 2 |
| | $\Delta$ | $m_t$ | $m_l$ | | 2 |
| | $\Lambda$ | $m_t \frac{(2m_l + m_t)}{(m_l + 2m_t)}$ | $\frac{(m_l + 2m_t)}{3}$ | $\frac{3m_l m_t}{(2m_l + m_t)}$ | 4 |
| $(111)/[\bar{2}11]/[0\bar{1}1]$ | $\Delta$ | $\frac{(2m_l + m_t)}{3}$ | $m_t$ | $\frac{3m_l m_t}{(2m_l + m_t)}$ | 2 |
| | $\Delta$ | $\frac{2}{3} m_t \frac{(2m_l + m_t)}{(m_l + m_t)}$ | $\frac{(m_l + m_t)}{2}$ | $\frac{3m_l m_t}{(2m_l + m_t)}$ | 4 |
| | $\Lambda$ | $m_t$ | $m_t$ | $m_l$ | 1 |
| | $\Lambda$ | $\frac{(8m_l + m_t)}{9}$ | $m_t$ | $\frac{9m_l m_t}{(8m_l + m_t)}$ | 1 |
| | $\Lambda$ | $\frac{m_t}{3} \frac{(8m_l + m_t)}{(2m_l + m_t)}$ | $\frac{(2m_l + m_t)}{3}$ | | 2 |
| $(110)/[001]/[0\bar{1}0]$ | $\Delta$ | $m_t$ | $\frac{(m_l + m_t)}{2}$ | $\frac{2m_l m_t}{(m_l + m_t)}$ | 4 |
| | $\Delta$ | $m_l$ | $m_t$ | $m_t$ | 2 |
| | $\Lambda$ | $\frac{(m_l + 2m_t)}{3}$ | $m_t$ | $\frac{3m_l m_t}{(m_l + 2m_t)}$ | 2 |
| | $\Lambda$ | $\frac{3m_l m_t}{(2m_l + m_t)}$ | $\frac{(2m_l + m_t)}{3}$ | $m_t$ | 2 |

**Table I.** Transport, width and confinement effective masses and subband degeneracies for three different technologically important semiconductor wafer orientations.

| Material | Valley | $m_l$ | $m_t$ |
| --- | --- | --- | --- |
| Silicon | Δ | 0.91 | 0.19 |
| | Λ | 1.7 | 0.12 |
| Germanium | Δ | 0.95 | 0.2 |
| | Λ | 1.64 | 0.08 |

**Table II.** Transverse and longitudinal effective masses for the Δ- and Λ-type valleys in silicon and germanium.